%% file: main.tex
\documentclass[11pt]{article}

\usepackage[preprint]{acl}

\usepackage{times}
\usepackage{latexsym}

\usepackage[T1]{fontenc}

\usepackage[utf8]{inputenc}

\usepackage{microtype}

\usepackage{inconsolata}

\usepackage{graphicx}

\usepackage{float}
\usepackage{placeins}

\usepackage{orcidlink}
\usepackage{amsmath}
\usepackage{amssymb}
\usepackage{tabularx} 
\usepackage{booktabs} 
\usepackage{xcolor}
\definecolor{zahracolor}{HTML}{006D77}   

\usepackage{tcolorbox}
\usepackage{enumitem}
\usepackage{dblfloatfix}

\input{macro/macro}

\title{Optimal Fusion Weights Form Intervals, Not Points: 

A Latency-Aware Study of Query-Adaptive WRRF}

\author{
  \textbf{Iris Ma\textsuperscript{1}}\,\orcidlink{0009-0003-3699-7981},
  \textbf{Zahra Shekarchi\textsuperscript{3}}\,\orcidlink{0009-0006-6166-1998},
  \textbf{Stuart Mcilroy\textsuperscript{3}}\,\orcidlink{0009-0004-2169-9831},
  \textbf{Mahtab Farrokh\textsuperscript{3}}\,\orcidlink{0009-0004-2493-5967},
\\
  \textbf{Fattane Zarrinkalam\textsuperscript{2,3}}\,\orcidlink{0000-0002-2102-9190},
  \textbf{Joel Stremmel\textsuperscript{3}}\,\orcidlink{0009-0002-4540-7982}
\\
\\
  \textsuperscript{1}University of California, Irvine, USA \quad
  \textsuperscript{2}University of Guelph, Canada \quad
  \textsuperscript{3}Thomson Reuters
\\
}

\begin{document}
\maketitle

\renewcommand{\thefootnote}{}\footnotetext{This work was conducted during the first author's internship at Thomson Reuters. \textcopyright\ 2026 Thomson Reuters. Licensed under CC BY-NC-ND 4.0.}
\renewcommand{\thefootnote}{\arabic{footnote}}

\input{sections/00-abstract}

\input{sections/01-introduction}

\input{sections/02-related-work}

\input{sections/03-method}
\input{sections/04-experiment}
\input{sections/05-results}

\input{sections/06-framework}
\input{sections/07-conclusion}
\input{sections/08-limitations}

\bibliography{main}

\appendix
\input{sections/09-appendix}

\end{document}

%% file: macro/macro.tex
\iftrue

\fi


%% file: sections/00-abstract.tex
\begin{abstract}

Reciprocal Rank Fusion (RRF) is widely used to combine sparse and dense retrievers in hybrid 
information retrieval, but a critical limitation is that it assigns a fixed, 
uniform weight to each retriever regardless of the query. Query-aware methods treat 
the per-query optimal weight as a single point, an assumption whose empirical validity 
has not been examined. We present an empirical study of eight dynamic fusion weight prediction strategies
across four datasets and four sparse--dense retriever pairs. 
When we compare these approaches to static fusion and a per-query oracle 
that bounds achievable performance, several findings emerge. First, the per-query optimal fusion weight is
typically an interval rather than a point. Queries can have multiple optimal intervals, and these can take on a variety of shapes. 
Second, 74 of 128 dynamic fusion methods tested were significantly better than standard RRF after false discovery rate correction. Third, dynamic fusion methods exhibit differing quality and latency characteristics that should be considered when choosing an appropriate fusion method. To address this last finding, we present a tiered decision framework that maps 
practitioner constraints, dataset properties, label availability, hardware, and latency 
budget to an appropriate method class. We release the code for our experiments at \url{https://github.com/thomsonreuters/query-based-rrf}.


%

\end{abstract}

%% file: sections/01-introduction.tex
\section{Introduction}
\label{sec:intro}

Hybrid information retrieval (IR) systems, which combine lexical signals from sparse retrievers 
with semantic embeddings from dense retrievers, have become the standard architecture 
for modern search applications~\cite{27ma2020hybrid,28berntson2023azure,04-sawarkar24blendedrag}.
To merge these disparate sources, Reciprocal Rank Fusion (RRF)~\cite{40selberg1997metacrawler,01-cormack09}
has been adopted in practice~\cite{31ding2020image,30jin2024matching,03-rackauckas2024rag,29ICLR2025_3640a199}. 
RRF is effective because it operates on ranks without training, bypassing score-scale mismatches between lexical and semantic retrievers~\cite{02-convex-bruch23}.

\input{figures/fig-intro-example}

However, standard RRF uses a fixed weight as a common default, 
treating all retrievers as equally valuable. Alternative score-based strategies, 
such as the convex combination of \citet{02-convex-bruch23}, 
fuse raw scores rather than ranks and require both per-retriever score normalization and a learned per-dataset weight.  
Consequently, this approach is less common in practice, since rank-based RRF needs neither. 
Weighted RRF (WRRF) makes the fusion weight tunable, and the natural follow-up is to choose it per query. 
We show this framing is itself wrong: optimal WRRF weights typically form intervals, 
not points.

\input{tables/tab-tiers}

Recent work has nonetheless pursued query-aware fusion weights, 
adjusting the WRRF parameter per query. 
Dynamic Alpha Tuning (DAT)~\cite{26hsu2025dat} uses a Large Language Model (LLM) to predict a query-specific fusion weight, 
exemplifying the query-aware weighting paradigm but introducing substantial inference latency. 
In addition, how much of the improvement above standard RRF dynamic methods actually capture remains an open question.

To assess this we evaluate a spectrum of fusion weight prediction strategies, ranging from a dataset-level constant and linear regression to fine-tuned encoders, in-context LLMs, and DAT, against a per-query oracle bounding maximum performance. We measure absolute retrieval quality, headroom recovery over standard RRF, and per-query inference latency.

Our contributions are as follows:
\begin{itemize}
    \item We characterize optimal fusion weights as a set-valued object, showing empirically that they typically form intervals rather than points. This clarifies when standard RRF is useful while exposing the limitations of framing weight prediction as point estimation.
    \item For alignment with the structure of optimal fusion weights as intervals, we use a satisficing loss to fine-tune small encoder-based weight predictors and interval-aware in-context learning for LLM-based weight prediction.
    \item We quantify the benefits of dynamic fusion while highlighting gaps between each method evaluated and a per-query oracle.  Our evaluations demonstrate statistically significant performance gains even after false discovery rate correction in 74 of 128 settings tested.
    \item We present a tiered decision framework (Table~\ref{tab:tiers}) that organizes fusion weight-prediction strategies along a measured latency spectrum, mapping practitioner constraints (labels, GPU capacity, latency budgets) to actionable method choices.
\end{itemize}

%% file: figures/fig-intro-example.tex
\begin{figure}[!tb]
\centering
\includegraphics[width=\columnwidth]{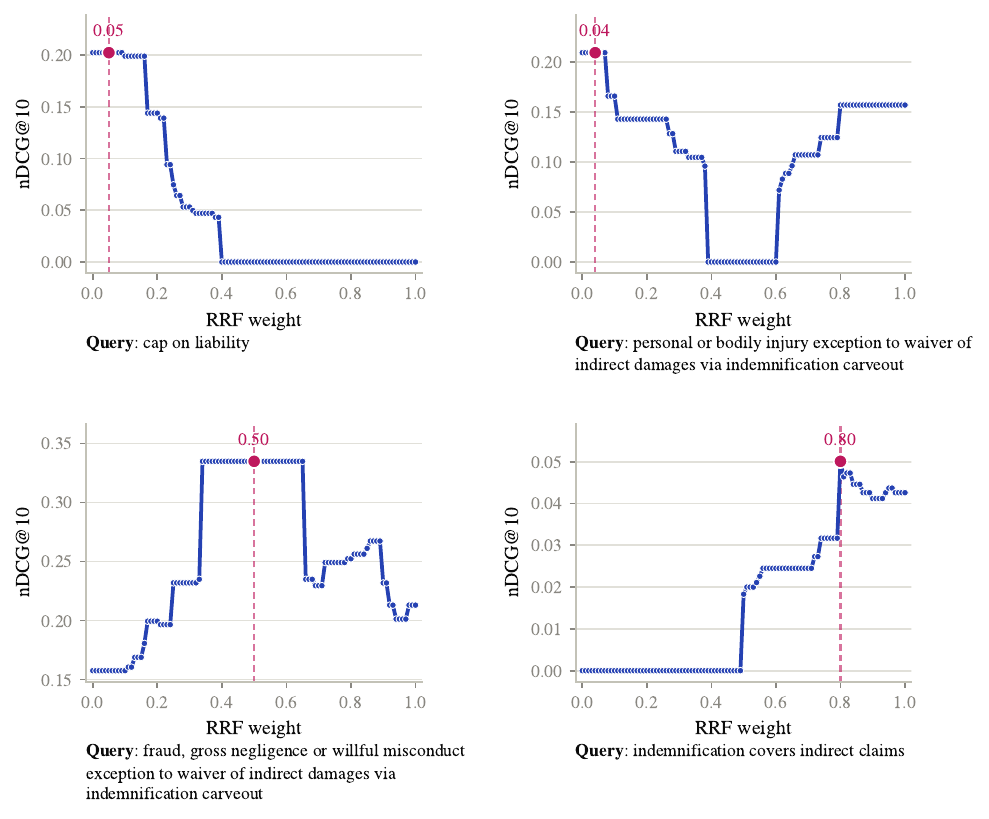}
\caption {Four ACORD queries with very different optimal RRF weights (more examples in Appendix~\ref{app:acord_contact_sheet}).}
\label{fig:intro_example}
\vspace{-10pt}
\end{figure}

%% file: tables/tab-tiers.tex
\begin{table*}[!tb]
\setlength{\tabcolsep}{3pt}
\renewcommand{\arraystretch}{1.15}
\centering
\resizebox{\textwidth}{!}{%
\begin{tabular}{llllllll}
\hline
\textbf{Tier} & \textbf{Sub-tier} & \textbf{Method} & \textbf{Latency ($95\%$ CI)} & \textbf{Infra} & \textbf{Training Data} & \textbf{Query-aware} & \textbf{Top-1 Retrieval-aware} \\
\hline
T0 & -- & RRF ($w=0.5$)         & $<1\,\mu$s                    & CPU                      & not required  & no & no \\
\hline
T1 & T1a  & Mean Optimal Weight & $<1\,\mu$s               & CPU                      & required      & no & no \\
   & T1b  & Linear Regression   & $265.00 \pm 3.00$\,µs          & CPU                      & required      & yes & no \\
\hline
T2 & T2a  & Small Encoder LM    & $5.24 \pm 0.04$\,ms    & GPU (self-host)          & required      & yes & no \\
   & T2b  & Small Encoder LM    & $8.46 \pm 0.02$\,ms    & GPU (self-host)          & required      & yes & yes \\
\hline
T3 & T3a  & DAT (zero-shot)     & $205.50 \pm 8.55$\,ms & GPU (self-host)/LLM API  & not required  & yes & yes \\
   & T3b  & Few-shot LLM        & $225.37 \pm 8.37$\,ms  & GPU (self-host)/LLM API  & required      & yes & no \\
   
\hline
\end{tabular}
} 
\caption{Three latency tiers motivated by inference latency clusters;
training requirements or data used for supervision define sub-tiers.
Latency reported as mean with 95\% CI over $n$ within-dataset means.}
\label{tab:tiers}
\vspace{-10pt}
\end{table*}

%% file: sections/02-related-work.tex
\section{Related Work}
\label{sec:related}

Multi-stage ranking systems rely on a retrieval phase and multiple re-ranking stages~\cite{02-convex-bruch23}.
Hybrid IR combines sparse lexical retrievers, such as BM25~\cite{33robertson1994okapi, 25-bm25-robertson},
which excel at exact keyword matching, with dense retrievers~\cite{34karpukhin2020dense, 35izacard2021unsupervised, 36chen2024m3},
which leverage transformer-based embeddings to capture semantic relationships.

To merge these retrieval streams, fusion methods have historically relied on static aggregation strategies.
Early approaches rooted in social choice theory, such as Borda Count~\cite{37aslam2001models} and round-robin 
interleaving~\cite{38etsioni1995multi}, often struggled with varying candidate pool sizes and ignored 
cross-system agreement. RRF~\cite{40selberg1997metacrawler, 01-cormack09} advanced retrieval fusion by 
applying a hyperbolic decay constant to reward consensus among retrievers. 

Still, standard RRF can be brittle. An optimized convex combination of normalized scores can perform 
better~\cite{02-convex-bruch23}, but such approaches remain limited because they rely on static, 
dataset-level weights.

The intuition that fusion weights should adapt to queries has driven the development
of dynamic strategies closely related to Query Performance Prediction (QPP), which aims to predict
retrieval effectiveness a priori~\cite{06-arabzadeh24qpp}.
Per-query weight optimization predates this line of work: \citet{53vogt1999fusion} derive a
closed-form optimal weight for linearly combining two retrieval systems, and
\citet{54sheldon2011lambdamerge} learn to weight multiple ranked lists per query, though for
merging query reformulations rather than sparse--dense retrievers.
Performance estimates can be translated into action in several ways.

In the context of fusion, a recent focus has been on using LLMs to calibrate retrieval parameters on the fly. 
AMSRAG~\cite{39dong2026adaptive} uses lightweight models such as ModernBERT~\cite{44warner-etal-2025-smarter} to classify queries into complexity tiers, 
then adjusts both fusion weights and the smoothing constant accordingly. 
Similarly, DAT~\cite{26hsu2025dat} uses an LLM to evaluate the top result for each retriever 
and assign a query-specific fusion weight.

While DAT improves retrieval quality over standard RRF, it treats fusion weight prediction as a point-estimation problem. Our in-context LLM approach contrasts with DAT in three key respects: we (1) fuse ranks rather than scores, bypassing score normalization; (2) condition the LLM on interval-valued few-shot examples rather than zero-shot prompts; and (3) rely, for runtime efficiency, on the query alone rather than candidate passages.

Beyond structural bounds, dynamic fusion introduces latency. Production pipelines keep early-stage retrieval lightweight to reserve compute budget for downstream LLM rankers~\cite{47sun2023chatgpt, 48tang-etal-2024-found} or LTR models~\cite{07-liu09ltr, 10-li24ltr}. Thus, while DAT-style LLM fusion improves query-aware weighting, its latency overhead is better allocated to downstream re-ranking.

We study several approaches for WRRF and compare against DAT. We chose to look at these WRRF variations 
because RRF is commonly used in practice and does not require standardized scores from the input retrievers. 
Inspired by DAT, we not only look at methods that are dataset-aware (setting the WRRF weight based on the 
mean optimal weight for a train split), but also consider methods that dynamically set the fusion weight 
for each query according to a query representation.

%% file: sections/03-method.tex
\section{Methodology}
\label{sec:method}

We first formalize Weighted Reciprocal Rank Fusion (WRRF) and construct the empirical optimal weight set $\mathcal{W}^*_q$. We then detail a spectrum of weight-prediction strategies, spanning dataset-level constants, discriminative encoders, and in-context LLMs, evaluated against this per-query oracle.

\subsection{Weighted Reciprocal Rank Fusion}

Given a query $q$, let $\mathcal{R}_{\text{sparse}}(q)$ and $\mathcal{R}_{\text{dense}}(q)$ denote the ranked document lists returned by the sparse and dense retrievers. For a document $d \in \mathcal{R}_{\text{sparse}}(q) \cup \mathcal{R}_{\text{dense}}(q)$, let $r_s(d|q)$ denote its rank in retriever $s \in \{\text{sparse}, \text{dense}\}$. 

We define the WRRF score for document $d$ given query $q$ as
\begin{equation}
\resizebox{0.9\linewidth}{!}{%
$\displaystyle \mathrm{WRRF}(d|q) = \frac{w_q}{k + r_{\text{sparse}}(d|q)} + \frac{1 - w_q}{k + r_{\text{dense}}(d|q)}$%
}
\end{equation}
where $w_q \in [0, 1]$ is the fusion weight assigned to the sparse retriever and $k=60$ is the standard smoothing constant~\cite{01-cormack09,02-convex-bruch23}. Standard RRF corresponds to $w_q = 0.5$; dynamic fusion aims to predict $w_q$ adaptively per query.

\subsection{Optimal Fusion Weights}
\label{subsec:optimal_weight_sets}

For each query, we identify its optimal weight set $\mathcal{W}^*_q \subseteq [0, 1]$ by sweeping $w \in [0,1]$ with step size $\Delta w = 0.01$ (a $10\times$ finer grid yields negligible differences; Appendix~\ref{app:grid-sensitivity}) and evaluating WRRF on top-$n$ candidate documents under metric $\mathrm{Metric}(q, w)$:
\begin{equation}
\resizebox{0.9\linewidth}{!}{%
$\displaystyle \mathcal{W}^*_q = \left\{ w \in [0,1] \;\middle|\; \mathrm{Metric}(q, w) = \max_{w' \in [0,1]} \mathrm{Metric}(q, w') \right\}$%
}
\end{equation}

The optimal set $\mathcal{W}^*_q$ generally forms a union of contiguous closed intervals $\mathcal{W}^*_q = \bigcup_i [a_i, b_i]$. The optimal set may yield: (i) a single point, (ii) a single contiguous interval, (iii) disjoint intervals, or (iv) $\mathcal{W}^*_q = \emptyset$ when no weight surfaces relevant documents (yielding no weight signal).


The fusion weight prediction strategies operate on the primary optimal interval $\mathcal{I}_q = [l_q, r_q]$, defined as the widest interval in $\mathcal{W}^*_q$. For point-estimation baselines requiring a scalar target $\bar{w}_q$, we take the midpoint $\bar{w}_q = \frac{l_q + r_q}{2}$ (breaking ties randomly).

\subsection{Evaluation Metrics}
\label{subsec:eval_metrics}

For each query, the oracle score is the maximum achievable metric value, $\mathrm{Oracle}(q) = \max_{w \in [0,1]} \mathrm{Metric}(q, w)$, attained by any weight in $\mathcal{W}^*_q$; averaged over the evaluation split, it upper-bounds what any predictor assigning a single weight per query can achieve. We define headroom as the gap between this oracle and standard RRF, $\mathrm{Headroom} = \mathrm{Oracle} - \mathrm{RRF}$, and each method's recovery as the fraction of that headroom it captures, $\mathrm{Recovery} = (\mathrm{Method} - \mathrm{RRF}) / (\mathrm{Oracle} - \mathrm{RRF})$.

\subsection{Significance Testing}
\label{subsec:stat-testing}

We assess each method's improvement over standard RRF using paired two-sided Student's $t$-tests over per-query metric scores~\cite{50urbano-stats}. Across the resulting comparisons, we apply Benjamini–-Hochberg False Discovery Rate (FDR) corrections~\cite{52-benjamini1995controlling} (Appendix~\ref{app:mtc}). For latency, we report mean runtimes with 95\% $t$-confidence intervals computed across sub-tier and tier evaluations.

\subsection{Weight Prediction Strategies}
\label{subsec:prediction}

To evaluate how effectively practical systems estimate fusion weights for unseen queries, we compare three families of strategies along an adaptivity--cost spectrum. We treat the trade-off between inference latency, training data dependencies, and retrieval accuracy as a central axis of our study.

\subsubsection{Dataset-Level Constant}
\label{subsubsec:global_mean}

As a corpus-calibrated static alternative to standard RRF, we set $w_q$ to the dataset-level mean $\bar{w}_{\text{corpus}} = \mathbb{E}_{q}[\bar{w}_q]$ computed over training queries. This baseline incurs zero inference overhead and stores only a single global scalar per dataset pair.

\subsubsection{Discriminative Query Predictors}
\label{subsubsec:discriminative}

To capture query-level variance, we train discriminative models mapping query $q$ (and optionally top retrieved passages) to weight $\hat{w}_q \in [0, 1]$. We evaluate three backbones of increasing capacity:
\begin{itemize}[leftmargin=*,noitemsep,topsep=2pt]
    \item \textbf{Linear Regression}~\cite{12-Hoerl01021970}: Ridge regression over TF-IDF query features.
    \item \textbf{RoBERTa}~\cite{14-liu2019roberta}: Encoder baseline fine-tuned on query embeddings.
    \item \textbf{ModernBERT}~\cite{44warner-etal-2025-smarter}: Fine-tuned in query-only and passage-conditioned variants.
\end{itemize}

We evaluate two design choices on ModernBERT:

\paragraph{Target form: point vs.\ interval.} Point targets ($\bar{w}_q$) penalize predictions that fall within valid optimal regions away from the midpoint. We contrast this against interval targets $\mathcal{I}_q = [l_q, r_q]$.

\paragraph{Loss form: MSE vs.\ satisficing.} For point targets, we optimize mean squared error. For interval targets, we apply a satisficing loss built on the $\epsilon$-insensitive tube of SVR~\cite{55smola2004tutorial}: rather than a fixed, symmetric $\epsilon$ around a scalar target, the tolerance zone is the query's own empirical interval $[l_q, r_q]$, so any prediction the data already certifies as optimal incurs no penalty~\cite{46simon1955behavioral}:
\begin{equation}
\resizebox{0.9\linewidth}{!}{%
$\displaystyle \mathcal{L}_{\text{sat}}(q) = \max(0, l_q - \hat{w}_q)^2 + \max(0, \hat{w}_q - r_q)^2$%
}
\end{equation}
which incurs zero loss when $\hat{w}_q \in [l_q, r_q]$ and penalizes deviations only outside interval boundaries.

\subsubsection{In-Context LLM Prediction}
\label{subsubsec:llm_icl}

We formulate dynamic weight prediction as a two-stage in-context learning (ICL) pipeline.

\paragraph{Exemplar retrieval.}
For a query $q$, we retrieve top-$k$ nearest neighbors from training queries using sparse lexical and dense embedding indices using a modified KATE setup~\cite{49liu-etal-2022-makes}. Deduplicating these lists produces an exemplar set capturing lexical and semantic query similarity.

\paragraph{Prompting and inference.}
We construct an ICL prompt pairing each retrieved neighbor $q_{\text{nbr}}$ with its optimal interval $\mathcal{I}_{q_{\text{nbr}}}$, instructing the LLM to predict scalar weight $\hat{w}_q \in [0, 1]$ for $q$. Presenting full interval boundaries rather than scalar point summaries preserves critical tolerance and range signals. We evaluate across model scales to test capacity dependencies and compare against DAT under matched backbones.

%% file: sections/04-experiment.tex
\section{Experiments}
\label{sec:experiment}

\subsection{Datasets and Metrics}

We evaluate on four datasets covering different domains: MS~MARCO~\cite{17-nguyen2016msmarco} (web search; dev split), NQ~\cite{18-kwiatkowski2019nq} (Wikipedia QA; dev split), NFCorpus~\cite{19-boteva2016nfcorpus} (medical IR; test split), and ACORD~\cite{20-wang2025acord} (legal contract retrieval; test split). Training splits are used to fit predictors and label optimal weight sets. We report MRR@10 for binary-labeled datasets for which queries typically only have one or two gold documents (MS~MARCO, NQ) and nDCG@10 for graded multi-document relevance (NFCorpus, ACORD).

\subsection{Retrievers}
\label{subsec:retrievers}

We pair two sparse and two dense retrievers into four hybrid configurations:
\begin{itemize}
    \item \textbf{Sparse:} BM25~\cite{25-bm25-robertson} (via Pyserini~\cite{21-Lin_etal_SIGIR2021_Pyserini}) and RM3~\cite{42abdul2004umass,41lavrenko2017relevance} (BM25 pseudo-relevance feedback query expansion).
    \item \textbf{Dense:} all-MiniLM-L6-v2~\cite{22-all-MiniLM-L6-v2} (lightweight bi-encoder) and Qwen3-Embedding-8B~\cite{43qwen3embedding} (LLM-based embedder).
\end{itemize}
We name each hybrid pair by concatenating a sparse and a dense abbreviation: BM (BM25) or RM (RM3), with the dense retriever appended in parentheses as (M) (all-MiniLM-L6-v2) or (Q) (Qwen3-Embedding-8B). For example, BM(Q) denotes BM25 and Qwen3-Embedding-8B, giving four configurations in total: BM(M), BM(Q), RM(M), RM(Q). Each retriever retrieves top-200 candidates before fusion (increasing the cutoff to 400 yielded no empirical gain).

\subsection{Predictors and Comparators}
\label{subsec:predictors}

\paragraph{Discriminative predictors.} We evaluate linear regression, RoBERTa-large, and ModernBERT-large, fine-tuned end-to-end (Appendix~\ref{app:training}).

\paragraph{In-context LLMs \& Comparators.} Our ICL pipeline evaluates Qwen3-32B~\cite{51yang2025qwen3} and Ministral-3-14B-Instruct-2512~\cite{45liu2026ministral}. Exemplars ($k=5$ sparse, $k=5$ dense) are retrieved over training queries as detailed in Appendix~\ref{app:llm}. We evaluate DAT under these identical LLM backbones.

\subsection{Latency Measurement}
\label{subsec:latency}

Per-query inference latency is benchmarked on an AWS \texttt{ml.g6e.2xlarge} instance (48\,GB VRAM), with LLM calls routed via Amazon Bedrock (timed full round-trip). Benchmarks use batch size 1, no local caching, and a warm-up pass over full evaluation splits. 


%% file: sections/05-results.tex
\section{Results}
\label{sec:results}

We organize the results around three parts. \S\ref{subsec:headroom} quantifies how much room exists above standard RRF for any per-query weight predictor, 
and how much of that room current methods recover. 
\S\ref{subsec:decompose} characterizes the structure of the per-query optimal weight set $\mathcal{W}^*_q$ and shows that the same characterization explains where the unrecovered headroom resides. 
\S\ref{subsec:distribution} reports the per-query optimal weight distribution and cross-dataset patterns. 

\subsection{Headroom and Recovery}
\label{subsec:headroom}

Table~\ref{tab:performance-comparison-full} reports the oracle, headroom, and each method's performance.
The headroom is largest on MS~MARCO and NQ (0.15--0.22),
smaller on ACORD and NFCorpus (0.05--0.09).
Translating these absolute differences into a comparable scale, Appendix~\ref{app:recovery-full} reports each method's recovery.

\input{tables/tab-performance-comparison-full}

Three observations stand out. First, recovery depends heavily on dense retriever capacity: when paired with the high-capacity embedder (Qwen3), several methods achieve 25--40\% recovery, peaking at 61.1\% (Ministral3 FS on NQ). Conversely, under the lightweight dense embedder (MiniLM), recovery drops sharply and frequently turns negative on out-of-domain datasets. Across BM25 pairings on MS~MARCO and NQ, top recoveries (>25\%) occur exclusively under high-capacity dense fusion. Second, median recovery across all setups is modest (12.2\%). Third, headroom and recovery are decoupled: configurations with the largest raw headroom, MS~MARCO and NQ under lightweight dense fusion, yield at most 18.8\% recovery.

\paragraph{Statistical significance.} Paired $t$-tests in Table~\ref{tab:performance-comparison-full} reveal a clear pattern: dynamic methods significantly outperform standard RRF in nearly all configurations on MS~MARCO (31 of 32 cells) and NQ (30 of 32), but far less frequently on NFCorpus (10 of 32) and ACORD (4 of 32). This trend directly mirrors headroom scale and training sample size. Significance is most detectable where headroom is large and query counts are high (MS~MARCO: 6,980; NQ: 2,893), and hardest to establish where both are small (ACORD: 57).

Adjusting for multiple comparisons across the 128 tests, 74 of the 75 raw significant cells survive Benjamini--Hochberg FDR control ($q \le .05$). Testing against the stronger Mean Optimal Weight baseline (MOW) rather than standard RRF yields 48 raw significant improvements, with 46 surviving FDR (details in Appendix~\ref{app:mtc}).

\paragraph{Extension to 3 retrievers.} we explore an extension to three retrievers(BM25, RM3, Qwen3) and report the details in Appendix~\ref{app:k3}. Extending fusion to more retrievers elevates the oracle ceiling across all datasets. Both evaluated predictors, ModernBERT Passage Cond.\ and Ministral3 FS (IW), consistently outperform unweighted three-way RRF across all datasets. These gains reach statistical significance ($p < .001$, surviving Bonferroni correction) on MS~MARCO and NQ for both models, and on NFCorpus for Ministral3 FS (IW).

\input{tables/tab-eval-interval-types}

\subsection{Anatomy of the Optimal Weight Sets}
\label{subsec:decompose}

The recovery numbers above represent statistics over heterogeneous queries. To understand where headroom resides and why it is difficult to capture, we partition each query's optimal weight set $\mathcal{W}^*_q$ into four categories (Table~\ref{tab:eval-interval-types}): empty, single point, single interval, and multi-interval.

\paragraph{Unreachable queries ($\mathcal{W}^*_q = \emptyset$).}
For 18--30\% of evaluation queries on MS~MARCO, NQ, and NFCorpus, no weight in $[0,1]$ surfaces a relevant document within the top 10. These queries score zero across all methods; while they do not affect relative recovery percentages, they cap the absolute oracle ceiling. On MS~MARCO under lightweight dense fusion (BM(M)), for instance, $30.2\%$ of queries yield zero metrics across all weights, capping the theoretical oracle at $.698$ ($1 - .302$). Against this bound, the realized oracle of $.443$ corresponds to an average MRR@10 of $\sim.63$ on reachable queries.

A detailed breakdown in Appendix~\ref{app:unreachable-breakdown} reveals two underlying drivers for unreachable queries. For $9\%$ (NQ) to $34\%$ (NFCorpus) of queries, neither retriever surfaces a relevant document within the top 200. The remaining cases occur when a relevant document appears only at deep ranks (e.g., near rank 200) in a single retriever, making it impossible to pull into the fused top-10 regardless of the weight setting.

\input{figures/fig-contains-half}

\paragraph{RRF-easy queries ($0.5 \in \mathcal{I}_q$).}
For a subset of reachable queries, the standard RRF weight of $0.5$ already lies within the query's optimal interval $\mathcal{I}_q$; on these queries RRF attains oracle performance by construction, so a dynamic predictor can at best match it. As shown in Figure~\ref{fig:contains_half}, this subpopulation is largest on NQ (27.4--38.2\%, peaking on BM(Q)) and MS~MARCO (24.2--26.3\%), moderate on NFCorpus (15.4--24.7\%), and smallest on ACORD (10.3--17.2\%). This cohort partly accounts for standard RRF's resilience as a baseline; on the remaining queries, where $0.5 \notin \mathcal{I}_q$, dynamic weighting has genuine room to improve and delivers statistically significant gains over RRF.

\paragraph{Interval-valued optima dominate over single points.}
\label{subsec:interval-dominate}
Among reachable queries ($\mathcal{W}^*_q \neq \emptyset$), optimal weight sets are predominantly intervals rather than isolated points. On MS~MARCO and NQ, single-point optima occur in only $\sim1\%$ of queries, whereas over $65\%$ yield a contiguous interval and 1--2\% feature disjoint intervals. Conversely, 19--28\% of ACORD queries exhibit single-point optima, an artifact of its coarse metric landscape and small evaluation set. This interval dominance creates an objective misalignment when training predictors on scalar targets $\bar{w}_q$, as valid predictions within $\mathcal{I}_q$ are penalized. While our $\epsilon$-insensitive satisficing loss removes this penalty, it closes only a small fraction of the recovery gap (Table~\ref{tab:recovery}), indicating that the primary bottleneck is the query-level predictability of $\mathcal{I}_q$ rather than loss formulation. Interval width helps explain why: queries where $0.5 \in \mathcal{I}_q$ have intervals 4--11$\times$ wider on average than queries where it does not, so the queries a predictor must move away from standard RRF to win on are also the ones with the narrowest, hardest-to-hit targets.

\paragraph{Grid resolution.} Comparing the $\Delta w = 0.01$ used to a finer $\Delta w = 0.001$ grid, interval statistics are stable across all dataset--retriever-pair configurations: optimal-interval length changes by at most
$0.0068$, interval disagreement by at most $0.011$, and the retrieval metric at the mean
best weight by at most $0.0007$ (Appendix~\ref{app:grid-sensitivity}). The $0.01$ grid is
therefore not an artifact of resolution.

\paragraph{Why is prediction hard?} Correlating a well-/poorly-predicted label with 15
query features (length, term rarity, named entities, WH-type, WordNet ambiguity) shows no
feature explains more than $\sim$1\% of variance in misprediction on the two large
datasets (strongest $|r| = 0.07$ MS~MARCO, $0.09$ NQ); even query ambiguity, our a
priori hypothesis, shows essentially no correlation ($|r| \le 0.08$). Mispredicted queries
are not identifiable from surface properties, reinforcing that the bottleneck is intrinsic
query-level unpredictability (Appendix~\ref{app:error-analysis}).

\subsection{Per-Query Optimal Weight Distribution}
\label{subsec:distribution}

Figure~\ref{fig:mean_optimal_weight} reports the distribution of per-query mean optimal weights for BM(Q) over reachable queries ($\mathcal{W}^*_q \neq \emptyset$); full distributions in Appendix~\ref{app:mean_optimal_weight_full}.

On MS~MARCO and NQ, distributions heavily skew toward low values of $w$ ($w \to 0$, favoring the dense retriever). Because both benchmarks are in-domain for the evaluated dense retrievers, the dense channel consistently dominates. This skew explains why the dataset-level MOW recovers a comparable fraction of headroom to complex predictors: when optimal weights are low, a single low constant closely approximates per-query optima.

\input{figures/fig-mean-optimal-weight}

On ACORD (out-of-domain legal IR), no systematic skew appears. Instead, optimal weights spread across $[0,1]$ with concentrations at the endpoints: 17--25\% of queries favor dense-only ($w = 0$) and 8--12\% favor sparse-only ($w = 1$), reflecting queries where single-channel retrieval strictly outperforms hybrid fusion. On NFCorpus under BM(Q), 15--25\% of queries cluster near $w = 0$ or $w = 0.5$, revealing distinct dense-only and balanced-fusion query cohorts; under BM(M), this mass spreads more uniformly across the interval.

%% file: tables/tab-performance-comparison-full.tex
\begin{table*}[!t]
\setlength{\tabcolsep}{3pt}
\renewcommand{\arraystretch}{1.1}
\centering
\resizebox{\textwidth}{!}{%
\begin{tabular}{l|cccc|c|cccc|c|cccc|c|cccc|c}
\hline
\textbf{Method} &
\multicolumn{5}{c|}{\textbf{MS~MARCO (MRR@10)}} &
\multicolumn{5}{c|}{\textbf{NQ (MRR@10)}} &
\multicolumn{5}{c|}{\textbf{ACORD (nDCG@10)}} &
\multicolumn{5}{c}{\textbf{NFCorpus (nDCG@10)}} \\
\cline{2-21}
 & \textbf{BM(M)} & \textbf{BM(Q)} & \textbf{RM(M)} & \textbf{RM(Q)} & \textbf{Avg}
 & \textbf{BM(M)} & \textbf{BM(Q)} & \textbf{RM(M)} & \textbf{RM(Q)} & \textbf{Avg}
 & \textbf{BM(M)} & \textbf{BM(Q)} & \textbf{RM(M)} & \textbf{RM(Q)} & \textbf{Avg}
 & \textbf{BM(M)} & \textbf{BM(Q)} & \textbf{RM(M)} & \textbf{RM(Q)} & \textbf{Avg} \\
\hline
Oracle (per-query best $w$) & .443 & .457 & .437 & .453 & .448
                            & .525 & .549 & .484 & .531 & .522
                            & .203 & .221 & .204 & .231 & .215
                            & .405 & .455 & .423 & .473 & .439 \\
Headroom (Oracle $-$ RRF)   & .157 & .166 & .163 & .175 & .165
                            & .147 & .157 & .186 & .220 & .177
                            & .071 & .083 & .078 & .088 & .080
                            & .054 & .068 & .068 & .076 & .066 \\
\hline
Standard RRF                & .286 & .292 & .274 & .278 & .282
                            & .378 & .391 & .298 & .312 & .345
                            & .132 & .138 & .126 & .143 & .135
                            & \textbf{.351} & .387 & .356 & .397 & .373 \\
Mean Optimal Weight         & .295$^{*}$ & .314$^{*}$ & .289$^{*}$ & .309$^{*}$ & .302
                            & .381 & .417$^{*}$ & .332$^{*}$ & .378$^{*}$ & .377
                            & .132 & .137 & .132 & .143 & .136
                            & .348 & .397$^{*}$ & .355 & .403$^{*}$ & .376 \\
Ridge Regressor (MBW)       & .296$^{*}$ & .317$^{*}$ & .289$^{*}$ & .310$^{*}$ & .303
                            & .383 & .416$^{*}$ & .330$^{*}$ & .377$^{*}$ & .377
                            & .138 & .137 & .130 & .145 & .137
                            & .349 & .397$^{*}$ & .356 & .402 & .376 \\
RoBERTa Regression (IW)     & .300$^{*}$ & .321$^{*}$ & .295$^{*}$ & .312$^{*}$ & .307
                            & .387$^{*}$ & .418$^{*}$ & .333$^{*}$ & .384$^{*}$ & .380
                            & .126 & .154 & .125 & .154 & .140
                            & .350 & .398$^{*}$ & .356 & .402 & .376 \\
ModernBERT (IW)             & \textbf{.305}$^{*}$ & .326$^{*}$ & .297$^{*}$ & .318$^{*}$ & .311
                            & .387$^{*}$ & .423$^{*}$ & .333$^{*}$ & .393$^{*}$ & .384
                            & .125 & \textbf{.156} & .125 & .155 & .140
                            & .350 & .399$^{*}$ & \textbf{.360} & .405$^{*}$ & .378 \\
ModernBERT Passage Cond.    & .304$^{*}$ & .327$^{*}$ & .296$^{*}$ & .325$^{*}$ & .313
                            & \textbf{.406}$^{*}$ & .426$^{*}$ & .341$^{*}$ & .397$^{*}$ & .392
                            & \textbf{.143} & .155$^{*}$ & .135 & \textbf{.159}$^{*}$ & \textbf{.148}
                            & .349 & \textbf{.407}$^{*}$ & .359 & .403$^{*}$ & \textbf{.379} \\
DAT Qwen3                   & .290$^{*}$ & .302$^{*}$ & .275 & .289$^{*}$ & .289
                            & .389$^{*}$ & .416$^{*}$ & .319$^{*}$ & .356$^{*}$ & .370
                            & .131 & .141 & \textbf{.145}$^{*}$ & .156$^{*}$ & .143
                            & .346 & .401$^{*}$ & .354 & .408$^{*}$ & .377 \\
DAT Ministral3              & .295$^{*}$ & .316$^{*}$ & .285$^{*}$ & .309$^{*}$ & .302
                            & .400$^{*}$ & .432$^{*}$ & .352$^{*}$ & .400$^{*}$ & .396
                            & .126 & .147 & .132 & .151 & .139
                            & .337 & .395 & .339 & \textbf{.409} & .370 \\
Qwen3 FS (IW)               & .304$^{*}$ & .335$^{*}$ & .295$^{*}$ & .328$^{*}$ & .315
                            & .391$^{*}$ & .433$^{*}$ & .347$^{*}$ & .411$^{*}$ & .396
                            & .137 & .143 & .126 & .142 & .137
                            & \textbf{.351} & .403$^{*}$ & .356 & .406 & .379 \\
Ministral3 FS (IW)          & .303$^{*}$ & \textbf{.350}$^{*}$ & \textbf{.299}$^{*}$ & \textbf{.346}$^{*}$ & \textbf{.325}
                            & .386 & \textbf{.447}$^{*}$ & \textbf{.371}$^{*}$ & \textbf{.446}$^{*}$ & \textbf{.412}
                            & .139 & .151 & .132 & .157 & .145
                            & .349 & .404$^{*}$ & .347 & .400 & .375 \\
\hline
\end{tabular}}
\caption{Per-query performance across all four retriever pairs, with the per-dataset average (\textbf{Avg}). $^{*}$ marks the method is significantly better than standard RRF ($p{<}.05$). Best value per column in \textbf{bold}.}
\label{tab:performance-comparison-full}
\vspace{-10pt}
\end{table*}

%% file: tables/tab-eval-interval-types.tex
\begin{table}[!tb]
\setlength{\tabcolsep}{3pt}
\renewcommand{\arraystretch}{1.1}
\centering
\resizebox{0.48\textwidth}{!}{%
\begin{tabular}{ll|cccc}
\hline
\textbf{Dataset} & \textbf{Retrievers} & \textbf{Empty} & \textbf{Single Point} & \textbf{Single Interval} & \textbf{Multi-Interval} \\
\hline
MS~MARCO  & BM(M) & 30.2 & 0.7 & 67.1 & 2.0 \\
          & BM(Q)  & 26.2 & 0.7 & 70.9 & 2.1 \\
          & RM(M) & 30.3 & 1.0 & 66.9 & 1.8 \\
          & RM(Q)  & 26.7 & 0.7 & 70.5 & 2.0 \\
\hline
NQ        & BM(M) & 27.1 & 0.5 & 71.0 & 1.3 \\
          & BM(Q)  & 20.4 & 0.7 & 77.3 & 1.7 \\
          & RM(M) & 29.7 & 1.0 & 68.5 & 0.9 \\
          & RM(Q)  & 21.5 & 0.4 & 77.0 & 1.0 \\
\hline
ACORD     & BM(M) & 0.0  & 21.1 & 78.9 & 0.0 \\
          & BM(Q)  & 0.0  & 19.3 & 77.2 & 3.5 \\
          & RM(M) & 0.0  & 28.1 & 71.9 & 0.0 \\
          & RM(Q)  & 0.0  & 26.3 & 70.2 & 3.5 \\
\hline
NFCorpus  & BM(M) & 22.6 & 6.2 & 70.3 & 0.9 \\
          & BM(Q)  & 18.9 & 5.3 & 74.9 & 0.9 \\
          & RM(M) & 23.2 & 7.1 & 69.0 & 0.6 \\
          & RM(Q)  & 18.3 & 3.4 & 77.1 & 1.2 \\
\hline
\end{tabular}}
\caption{Structural breakdown (\%) of the per-query optimal weight set $\mathcal{W}^*_q$ on the evaluation split. Values represent percentages of total queries.}
\label{tab:eval-interval-types}
\vspace{-10pt}
\end{table}

%% file: figures/fig-contains-half.tex
\begin{figure}[!tb]
\centering
\includegraphics[width=0.5\textwidth]{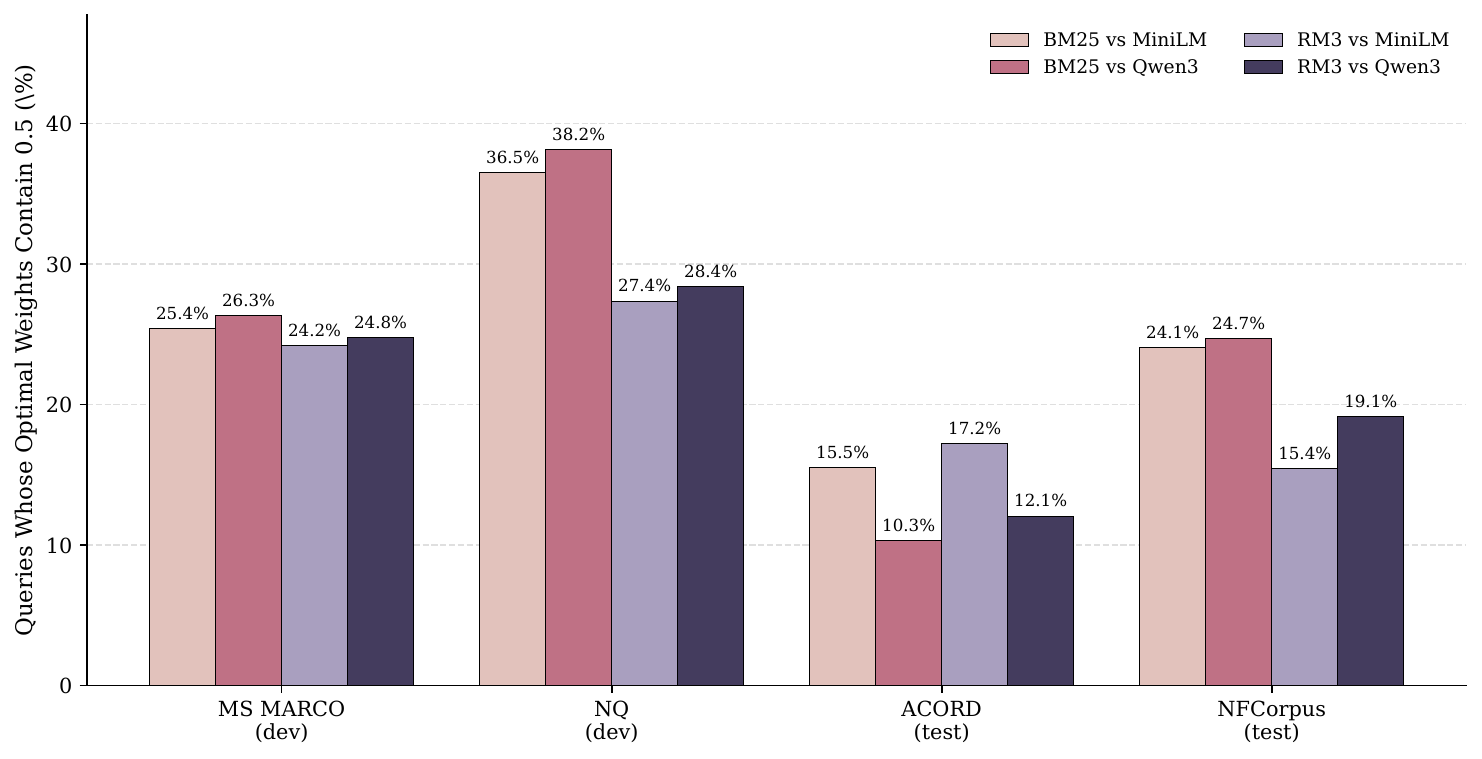}
\caption{Percentage of evaluation queries for which $0.5 \in \mathcal{I}_q$. On these queries RRF attains the per-query oracle by construction; a dynamic predictor that selects $\hat{w}_q \neq 0.5$ can only match or underperform RRF.}
\label{fig:contains_half}
\vspace{-10pt}
\end{figure}

%% file: figures/fig-mean-optimal-weight.tex
\begin{figure}[!tb]
\centering
\includegraphics[width=0.5\textwidth]{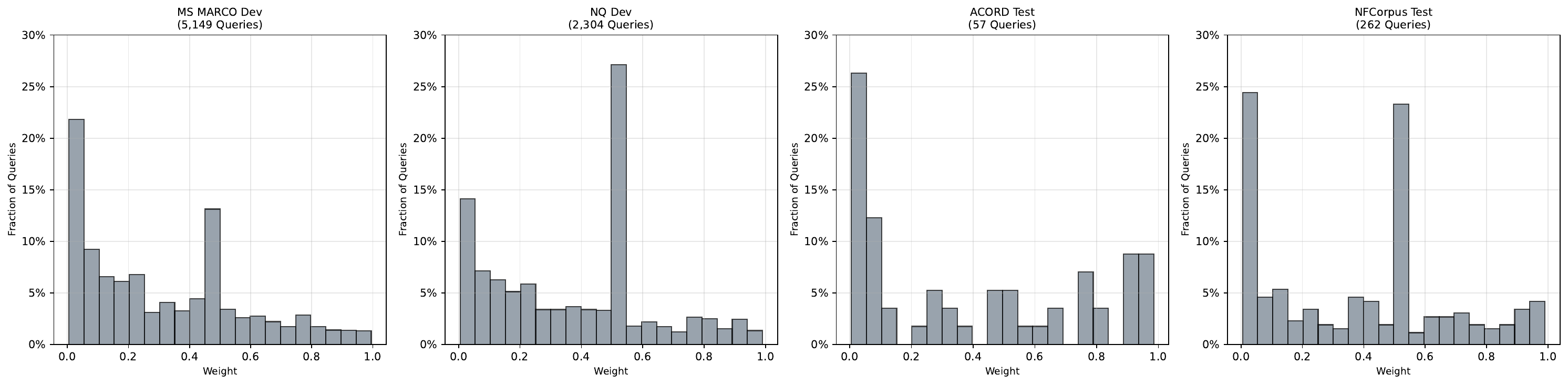}
\caption{Distribution of the per-query mean optimal weight for the BM25 and Qwen3 combination ($w{=}0$: dense only; $w{=}0.5$: RRF; $w{=}1$: sparse only). Unreachable queries omitted. Other pairs: Appendix~\ref{app:mean_optimal_weight_full}.}
\label{fig:mean_optimal_weight}
\vspace{-10pt}
\end{figure}

%% file: sections/06-framework.tex

\section{A Decision Framework for Fusion Method Selection}
\label{sec:framework}

\input{figures/fig-tradeoff}

Per-query inference latencies span sub-$\mu$s to $280$~ms and cluster into three regimes: 
Tier~0 and Tier~1 methods incur no per-query forward pass; 
Tier~2 invokes a fine-tuned encoder ($[5.9, 6.8]$~ms, $95\%$ CI);
Tier~3 invokes an LLM ($[209.1, 221.8]$~ms, $95\%$ CI).
Figure~\ref{fig:tradeoff} plots quality versus latency across datasets. 
We organize the discussion around three latency tiers observable in the figure, 
separating methods by inference runtime and corresponding infrastructure. 
As detailed in Table~\ref{tab:tiers}, sub-tiers within each tier share similar latency and infrastructure but differ in terms of training requirements or supervision data.

\subsection{Two Prerequisite Checks}
Selecting an appropriate dynamic fusion method depends on two practical criteria. 

\paragraph{Dataset sensitivity.} Evaluating relative headroom, $\frac{\text{Oracle} - \text{RRF}}{\text{Oracle}}$, reveals whether available headroom exceeds a threshold $\delta$ required to justify higher-tier computational overhead.

\paragraph{Training data availability.} Tiers T1, T2, and T3b all require training queries. Without labeled data, candidate choices naturally reduce to training-free baselines (T0) or zero-shot prompts (T3a).

\subsection{Tier Properties}
\textbf{T0} is appropriate when no labels and no LLM are available, 
or as a zero-latency method in hybrid retrieval. \textbf{T1} applies a learned weight on CPU: T1a captures a small but consistent gain on responsive datasets (details in Appendix~\ref{app:recovery-full}),
and T1b performs comparably to T1a. Thus, T1 is a suitable low-latency choice when no GPU is available or when a simple model is preferred for deployment. 
\textbf{T2} shares the same latency tier across sub-tiers ($6.313 \pm 0.446$\,ms per query), 
and on responsive datasets recovers much of the achievable gain at a fraction of the LLM inference latency. 
T2b is competitive across datasets, the strongest T2 method overall, 
and is a latency-effective choice when a GPU and top-1 passages are available at fusion time. 
\textbf{T3} invokes an LLM: T3a is the only method above T0 that requires no labeled training data, 
while T3b is the strongest method on NQ and MS~MARCO. 
Within-tier variation in T3 is substantial (see LLM rows of Table~\ref{tab:recovery}, Appendix~\ref{app:recovery-full}).
We sample at a low temperature ($0.1$), so run-to-run variation is minor. 
T3 fits best when fusion is the final stage and the latency budget allows the LLM call; otherwise, the same budget may serve a downstream re-ranker better. Non-retrieval-aware methods (T0--T2a, T3b) depend only on the query.

%% file: figures/fig-tradeoff.tex
\begin{figure*}[!tb]
\centering
\includegraphics[width=\textwidth]{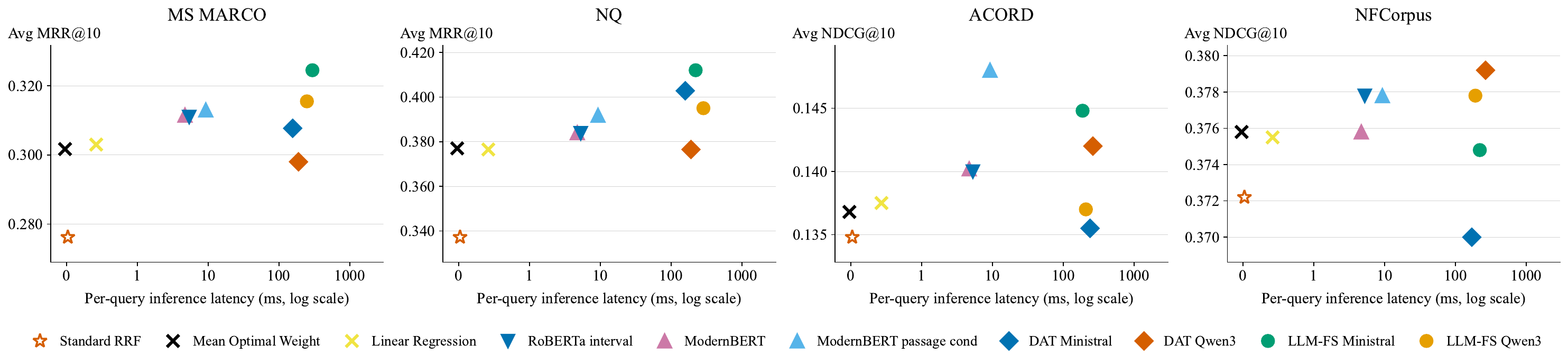}
\caption{Per-query latency vs.\ ranking quality across four datasets (MRR@10 for MS~MARCO and NQ, nDCG@10 for ACORD and NFCorpus). Each point averages over 44.4 retriever combinations per dataset, for each fusion method. }
\label{fig:tradeoff}
\vspace{-10pt}
\end{figure*}

%% file: sections/07-conclusion.tex
\section{Conclusion}
\label{sec:discussion}

Our work demonstrates that optimal fusion weights in hybrid retrieval are typically intervals, not points.
For a fraction of queries, standard 50-50 RRF weights already fall inside this interval,
which partly explains this strategies' resilience as a baseline. 

However, across hybrid retrieval settings tested, dynamic weight prediction improves significantly over standard RRF for most
MS~MARCO and NQ configurations (even after FDR correction). In these and other settings evaluated headroom relative to an oracle remains, even for LLM-based predictors that exhibit higher latency.  This
indicates that query-level predictability in addition to headroom availability, is an important area for future work on hybrid search.

Because of the latency, compute, and data labeling costs associated with some dynamic fusion methods, we organized the static and dynamic fusion methods tested in our analysis into a tiered decision 
framework.  The framework maps technical constraints 
to an appropriate method class, helping practitioners realize the potential value afforded by available fusion methods while respecting practical limitations.

%% file: sections/08-limitations.tex


\section*{Limitations}
\label{Limitations}
\begin{itemize}
\item We evaluate only on English benchmarks, where both retriever families are mature. The set-valued characterization should be language-agnostic, but the shape of the weight intervals themselves may shift in multilingual settings.
\item We sample LLM predictions at temperature 
0.1, so absolute LLM-method numbers may vary slightly across runs.
\end{itemize}

%% file: sections/09-appendix.tex
\section{Dataset Statistics}
\label{app:datasets}

Table~\ref{tab:dataset-stats} summarizes split sizes, total passage collection sizes, and qrels graded-relevance scales for all evaluation datasets (Section~\ref{sec:experiment}).

\input{tables/tab-dataset-stats}

\section{Training Details}
\label{app:training}

\subsection{Linear Regression}
\label{app:training-linreg}

We implement the linear regression baseline (Section~\ref{subsubsec:discriminative}) using scikit-learn~\cite{24-pedregosa2011scikit}. Queries are vectorized via \texttt{TfidfVectorizer} with 1--4 word n-grams (\texttt{ngram\_range=(1,4)}), English stop-word removal, sublinear term-frequency scaling (\texttt{sublinear\_tf=True}), and document-frequency cutoffs $\texttt{min\_df}=2$ and $\texttt{max\_df}=0.95$ over an unbounded vocabulary. Ridge regression (\texttt{Ridge}) is fit with regularization strength $\alpha = 100$, a learned intercept, and default solver selection. Predictions are clipped to $[0.01, 1.0]$ at inference time.

\subsection{RoBERTa}
\label{app:training-roberta}

We fine-tune \texttt{roberta-large} via Hugging Face Transformers Trainer, adding a linear regression head over the pooler output with $0.01$ dropout on attention and pooled representations. Queries are truncated to length $64$. Optimization uses AdamW (learning rate $2{\times}10^{-5}$, linear schedule, $500$ warmup steps, weight decay $0.001$, gradient clipping $0.5$) with per-device batch size $256$ and \texttt{fp16} mixed precision for up to $10$ epochs (seed $42$). We evaluate every $200$ steps and apply early stopping (patience $10$) based on evaluation loss. The interval variant minimizes satisficing loss $\mathcal{L}_{\text{sat}}$ against $[l_q, r_q]$; the point variant minimizes MSE against scalar target $\bar{w}_q$.

\subsection{ModernBERT}
\label{app:training-modernbert}

We fine-tune \texttt{answerdotai/ModernBERT-large} in two input configurations: a query-only variant (max length $64$, batch size $256$) mirroring RoBERTa, and a passage-conditioned variant (\texttt{[CLS] Query [SEP] Sparse Top-1 [SEP] Dense Top-1 [EOS]}, max length $1024$, batch size $8$) taking top-1 passages from the evaluation retrievers. Both attach a linear head over the \texttt{[CLS]} token (dropout $0.01$) and optimize $\mathcal{L}_{\text{sat}}$ against interval targets $[l_q, r_q]$. All optimization, scheduling, loss evaluation, early-stopping criteria, and random seeds match the RoBERTa setup (Appendix~\ref{app:training-roberta}).

\section{In-Context LLM Prediction Details}
\label{app:llm}

\subsection{Exemplar Retrieval}
\label{app:llm-retrieval}

For target query $q$, we retrieve and union top-$k$ ($k=5$) training queries from two independent indices built over the training set:
\paragraph{Dense index.} Queries are embedded via SentenceTransformers (\texttt{trust\_remote\_code=True}, \texttt{normalize\_embeddings=True}) in \texttt{bfloat16}. Target $q$ is embedded identically and scored by cosine similarity (equivalent to inner product post-normalization) to retrieve top $k=5$.
\paragraph{Sparse index.} Training query texts are indexed via \texttt{bm25s}\footnote{\url{https://github.com/xhluca/bm25s}} with default tokenization and BM25 parameters to retrieve top $k=5$.
\paragraph{Union and exemplar payload.} Deduplicating by \texttt{query\_id} yields up to 10 exemplars. Each exemplar carries its optimal weight interval $\mathcal{I}_{q_{\text{nbr}}}$ and its training-split metric score (MRR@10 or nDCG@10) under its best weight.

\subsection{LLM Generation}
\label{app:llm-generation}

We evaluate Qwen3-32B and Ministral-3-14B-Instruct-2512 using the Hugging Face library with \texttt{device\_map="auto"}. Inputs are processed using official chat templates (\texttt{add\_generation\_prompt=True}) alongside left-padded batched tokenization (\texttt{pad\_token = eos\_token}). The full prompt template is shown in Figure~\ref{fig:prompt_template}.

\paragraph{Qwen3.} Loaded via \texttt{AutoModelForCausalLM} with 4-bit quantization (\texttt{BitsAndBytesConfig}, \texttt{bnb\_4bit\_compute\_dtype=torch.bfloat16}). Reasoning traces are disabled (\texttt{enable\_thinking=False}). Generation uses \texttt{max\_new\_tokens=10}, \texttt{temperature=0.1}, \texttt{do\_sample=True}, and \texttt{pad\_token\_id=eos\_token\_id}.

\paragraph{Ministral-3.} Loaded in FP8 via \texttt{Mistral3ForConditionalGeneration} with \texttt{FineGrainedFP8Config(dequantize=True)}. Inputs are filtered to \texttt{input\_ids} and \texttt{attention\_mask}. Generation uses \texttt{max\_new\_tokens=10}, \texttt{temperature=0.1}, and \texttt{do\_sample=True}.

\paragraph{Invalid-output handling.} Output strings are parsed with regular expression \texttt{\textbackslash d*\textbackslash.\textbackslash d+} for floats, falling back to \texttt{\textbackslash b[01]\textbackslash b} for integers. Extracted values are clipped to $[0, 1]$ and rounded to two decimal places. Parse failures default to $\hat{w}_q = 0.50$.

\input{figures/fig-prompt-template}

\section{Per-Query nDCG Sensitivity to Fusion Weight}
\label{app:acord_contact_sheet}

Figure~\ref{fig:acord_contact_sheet} illustrates nDCG@10 across varying RRF fusion weights for nine representative ACORD queries (RM3, Qwen3). The dashed lines mark per-query optimal weights; flat plateaus around these peaks confirm that many queries exhibit optimal weight intervals rather than isolated points.

\input{figures/fig-acord-contact-sheet}

\section{Per-Query Optimal Weight Distribution: Full Results}
\label{app:mean_optimal_weight_full}

Figure~\ref{fig:mean_optimal_weight_full} provides the per-query mean optimal weight distributions on the evaluation split across all four retriever pairs.

\input{figures/fig-mean-optimal-weight-full}

\section{Additional Results: Recovery, Significance, and Three-Retriever Fusion}
\label{app:additional-results}

\subsection{Anatomy of Unreachable Queries}
\label{app:unreachable-breakdown}

Table~\ref{tab:unreachable-breakdown} categorizes unreachable queries ($\mathcal{W}^*_q = \emptyset$) by candidate availability prior to fusion. This isolates complete retrieval failures (absent from both top-200 candidate lists) from depth-ranking failures (present in candidate pools but too deeply ranked to surface in the fused top 10 under any weight $w \in [0,1]$).

\input{tables/tab-unreachable-breakdown}

\subsection{Interval Width Statistics}
\label{app:interval-width}

Table~\ref{tab:eval-interval-types} reports how often $\mathcal{W}^*_q$ is empty, a single point, a single interval, or multi-interval, but not how wide those intervals are -- the quantity that determines how hard a query is to hit. Table~\ref{tab:interval-width} reports the width of the primary optimal interval $\mathcal{I}_q$ over reachable queries, both overall and split by whether $0.5 \in \mathcal{I}_q$.

\input{tables/tab-interval-width}

\subsection{Full Headroom Recovery}
\label{app:recovery-full}

Table~\ref{tab:recovery} presents the per-cell headroom recovery scores, across all dataset and retriever combinations.

\input{tables/tab-recovery}

\subsection{Multiple-Testing Correction}
\label{app:mtc}

Table~\ref{tab:mtc} summarizes multiple-testing adjustments across the 128 paired $t$-tests described in Section~\ref{subsec:stat-testing} (8 query-adaptive methods $\times$ 16 dataset-retriever configurations), with Bonferroni and BH-FDR correction for the multiple-comparisons setting. We also evaluate whether a dataset-level static weight baseline (Mean Optimal Weight, MOW) independently outperforms standard RRF across its 16 configurations ($\alpha = 0.05/16 \approx 0.0031$). MOW nominally outperforms standard RRF in 9 configurations, 8 of which survive Bonferroni correction, primarily concentrated on Qwen3-paired configurations.

\input{tables/tab-mtc}

\subsection{Grid Discretization Sensitivity}
\label{app:grid-sensitivity}

We test whether the $\Delta w = 0.01$ grid (\S\ref{subsec:optimal_weight_sets}) is fine
enough to localize $\mathcal{W}^*_q$ by comparing it against a $10\times$ finer
$\Delta w = 0.001$ grid. For each query and retriever pair we measure (i) the change in
total optimal-interval length, (ii) interval disagreement (union $-$ intersection) between
the two grids, and (iii) the resulting change in the retrieval metric at the mean best
weight. Table~\ref{tab:grid-sensitivity} reports the range across the four retriever pairs
per dataset. The effect is negligible everywhere: interval length changes by at most
$0.0068$, disagreement by at most $0.011$, and the retrieval metric by at most $0.0007$,
consistent with retrieval metrics being piecewise constant in $w$, so a finer grid only
sharpens interval endpoints rather than creating or removing them. The $\Delta w = 0.01$
grid used throughout the paper is therefore not an artifact of the chosen resolution.

\input{tables/tab-grid-sensitivity}

\subsection{Error Analysis: Query-Feature Correlates of Misprediction}
\label{app:error-analysis}

To test whether mispredicted queries are identifiable from surface query properties, we
label each query as well- or poorly-predicted (predicted weight outside vs.\ inside
$\mathcal{W}^*_q$, pooled over methods) and compute point-biserial correlations against 15
query features, including length (words/characters), average term rarity, named-entity
count, digit presence, WH-question type, and WordNet-based ambiguity (polysemy).
Table~\ref{tab:error-analysis} reports the strongest correlate per dataset.

No query feature explains more than $\sim$1\% of variance in misprediction on the two
large datasets (strongest $|r| = 0.07$ on MS~MARCO, $0.09$ on NQ), and the
ambiguity signal shows essentially no correlation on any dataset ($|r| \le 0.08$). The one
directionally consistent, if weak, signal across all four datasets is that longer queries
with rarer terms are marginally harder to predict. Mispredicted queries are thus not
identifiable from surface query properties alone, reinforcing that the per-query optimum is
not predictable from the query text alone (\S\ref{subsec:decompose}).

\input{tables/tab-error-analysis}

\subsection{Extending to k=3}
\label{app:k3}

Fusing three retrievers (BM25, RM3, Qwen3) across the 3-retriever simplex raises the oracle performance ceiling across all datasets compared to 2-retriever pairings (Table~\ref{tab:k3-oracle}).

Table~\ref{tab:k3} evaluates Ministral3 FS (IW) and ModernBERT Passage Conditioned against unweighted 3-way RRF. Both dynamic methods consistently surpass unweighted 3-way RRF. Improvements are statistically significant ($p < .001$, Bonferroni-corrected) on MS~MARCO, NQ, and NFCorpus (ModernBERT PC only), but non-significant on ACORD.

\input{tables/tab-k3-oracle}
\input{tables/tab-k3}

\paragraph{Training-budget caveat.} ModernBERT-PC models reflect varying training budgets: NFCorpus (10 epochs), MS~MARCO and NQ (capped at 20,000 steps; $\approx$0.5 and $\approx$2.4 epochs), and ACORD (48-step sanity check). The ACORD and MS~MARCO/NQ ModernBERT-PC numbers are therefore not directly comparable to a converged run.

\section{Licenses and Terms of Use}
\label{app:licenses}

All datasets, models, and software used in this work are publicly released for research use and, to our knowledge, do not contain personally identifying information about private individuals or offensive content. Our use is consistent with their stated licenses and terms.

\begin{itemize}
    \item \textbf{Datasets}: MS~MARCO (Microsoft Non-Commercial Research License), Natural Questions (CC BY-SA 3.0), NFCorpus (Non-commercial research use), ACORD (CC BY 4.0).
    \item \textbf{Retrieval Software}: BM25 and RM3 via Pyserini (Apache 2.0); \texttt{bm25s} (MIT).
    \item \textbf{Models}: all-MiniLM-L6-v2, Qwen3-Embedding-8B, ModernBERT-large, Qwen3-32B, and Ministral-3-14B-Instruct-2512 (Apache 2.0); RoBERTa-large (MIT).
    \item \textbf{Libraries}: scikit-learn and PyTorch (BSD-3-Clause); Hugging Face Transformers, Datasets, and SentenceTransformers (Apache 2.0).
\end{itemize}

\section{AI Assistance Disclosure}
\label{app:ai-disclosure}

We used general-purpose AI assistants in a limited, supervised capacity: grammar and phrasing polish during writing, and brainstorming during research discussions.

%% file: tables/tab-dataset-stats.tex
\begin{table*}[t]
\centering
\setlength{\tabcolsep}{5pt}
\begin{tabular}{lrrrrl}
\toprule
\textbf{Dataset} & \textbf{\#Train} & \textbf{\#Dev} & \textbf{\#Test} & \textbf{\#Passages} & \textbf{Relevance scale} \\
\midrule
MS~MARCO & 502{,}939 & 6{,}980 &      43 & 8{,}841{,}823 & $\{0,1\}$ (train/dev); $\{0,1,2,3\}$ (test) \\
NQ       &  92{,}468 & 2{,}893 &       -- & 3{,}040{,}270 & $\{0,1\}$ (train); $\{0,1,2,3,4\}$ (dev) \\
ACORD    &        51 &       6 &      57 &     3{,}931 & $\{0,1,2,3,4\}$ \\
NFCorpus &   2{,}590 &     324 &     323 &     3{,}633 & $\{0,1,2\}$ \\
\bottomrule
\end{tabular}
\caption{Per-dataset statistics: number of queries in each split, total passage-collection size, and graded-relevance scale used in the qrels.}
\label{tab:dataset-stats}
\end{table*}

%% file: figures/fig-prompt-template.tex
\begin{figure*}[t]
\begin{tcolorbox}[
  colback=gray!5,
  colframe=gray!50,
  title=LLM Weight Prediction Prompt Template,
  center title,
  width=\textwidth,
  boxsep=4pt,
  fontupper=\small\ttfamily
]
\textbf{System:}\\
You are an expert search system optimizer. Your task is to predict the optimal weight for combining a sparse retriever and a dense retriever for a given search query.\\[6pt]

\textbf{STRICT CONSTRAINT:}\\
You must output ONLY a single float number between 0.00 and 1.00. Do not include any reasoning, markdown formatting, explanation, or conversational text. Just the number.\\[6pt]

\textbf{\#\#\# Context (Similar Queries from Train Split):}\\
For each historical query, you are provided its score and its optimal weights (a list of weights that yielded the best score, representing the weight given to the sparse retriever).\\
{[1]} Query: "[Example Query 1]" \quad Optimal Weights: [0.12, 0.25] \quad Score (nDCG@10): 0.842\\
\dots\\
{[10]} Query: "[Example Query 10]" \quad Optimal Weights: [0.75, 0.90] \quad Score (nDCG@10): 0.913\\[6pt]

\textbf{\#\#\# Task:}\\
Based on the optimal weights of the similar context queries, predict a single optimal weight for the Target Query.\\[6pt]

\textbf{\#\#\# Target Query:}\\
Query: \{\{target\_query\}\}\\
Predicted Optimal Weight:
\end{tcolorbox}
\caption{Few-shot prompt for LLM-based weight prediction.}
\label{fig:prompt_template}
\end{figure*}

%% file: figures/fig-acord-contact-sheet.tex
\begin{figure*}[t]
\centering
\includegraphics[width=\textwidth]{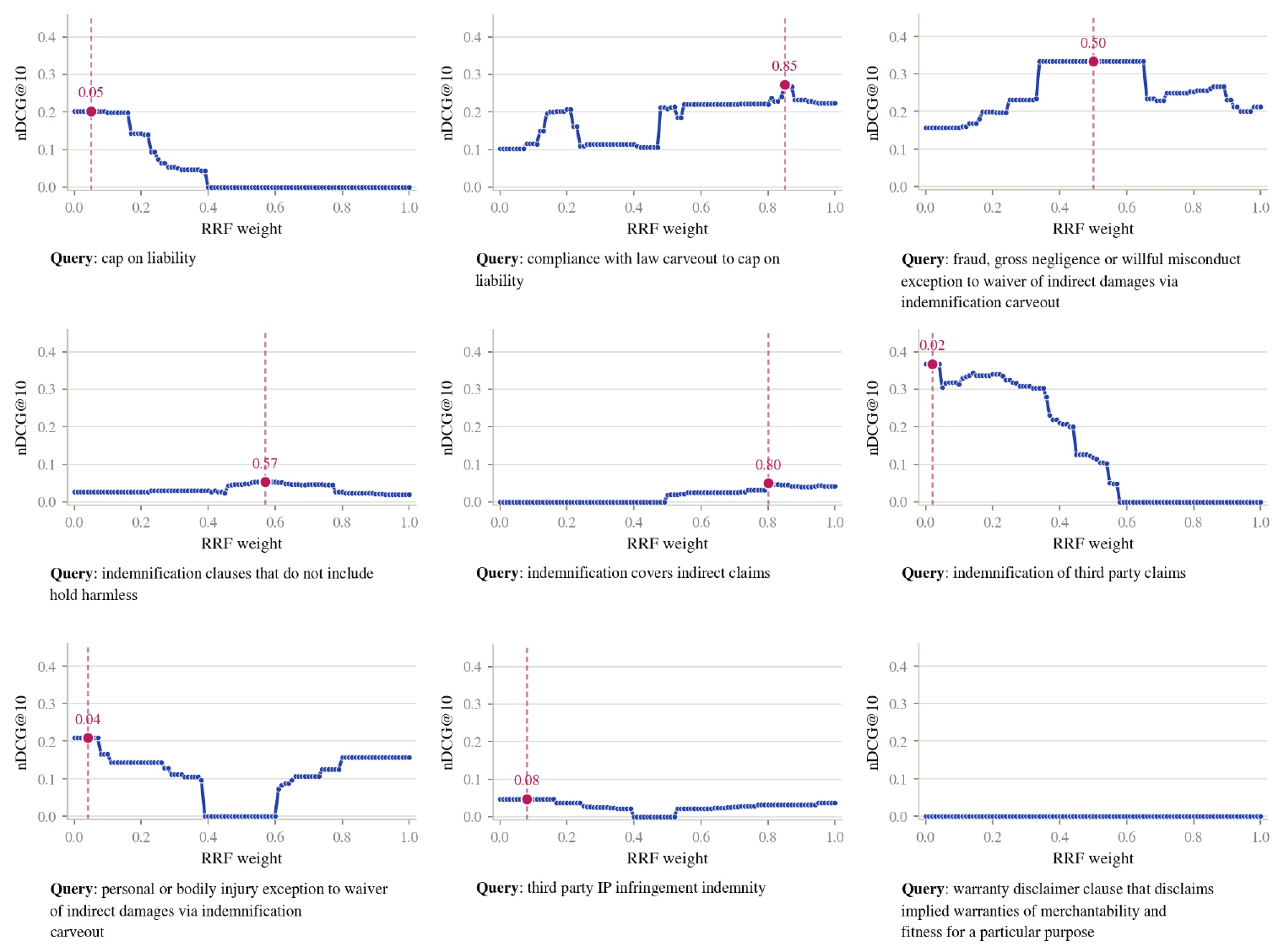}
\caption{nDCG@10 vs.\ RRF fusion weight for nine ACORD queries (RM3+Qwen3, train, top-200). Dashed line: each query's mean optimal weight.}
\label{fig:acord_contact_sheet}
\end{figure*}

%% file: figures/fig-mean-optimal-weight-full.tex
\begin{figure*}[!tb]
\centering
\includegraphics[width=\textwidth]{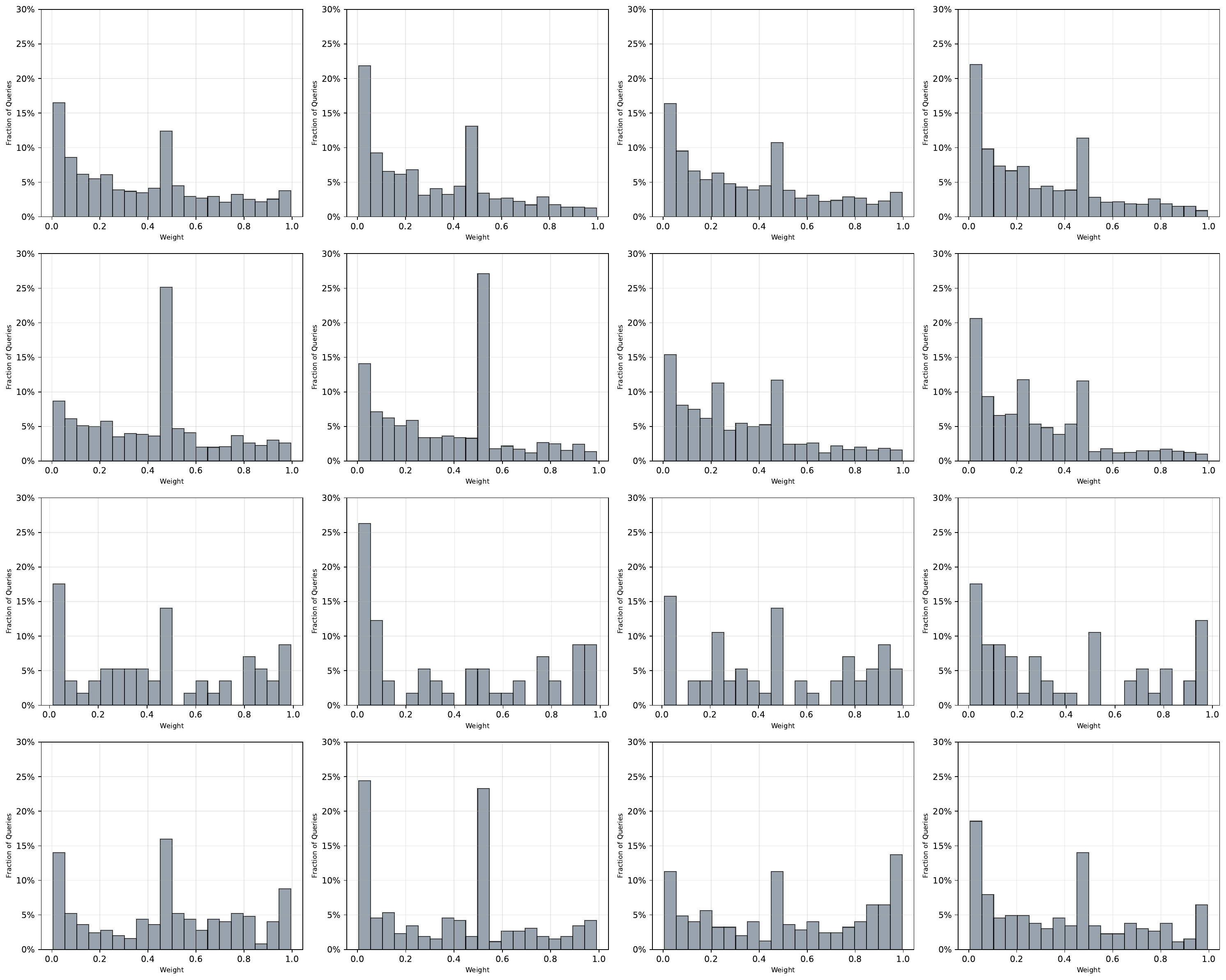}
\caption{Distribution of the per-query mean optimal weight on the evaluation split across all four retriever pairs, $\frac{1}{|\mathcal{I}_q|}\int_{\mathcal{I}_q} w\,\mathrm{d}w$ for queries with $\mathcal{W}^*_q \neq \emptyset$.}
\label{fig:mean_optimal_weight_full}
\end{figure*}

%% file: tables/tab-unreachable-breakdown.tex
\begin{table*}[!tb]
\centering
\small
\setlength{\tabcolsep}{12pt} 
\renewcommand{\arraystretch}{1.1}
\begin{tabular}{ll c cc}
\toprule
\textbf{Dataset} & \textbf{Retrievers} & \textbf{\# Unreachable} & \textbf{Neither in Top-200} & \textbf{$\ge$1 in Top-200} \\
\midrule
MS~MARCO  & BM(M) & 2106 & 20.8 & 79.2 \\
          & BM(Q) & 1831 & 12.9 & 87.1 \\
          & RM(M) & 2115 & 20.8 & 79.2 \\
          & RM(Q) & 1865 & 12.7 & 87.3 \\
\midrule
NQ        & BM(M) &  785 & 20.1 & 79.9 \\
          & BM(Q) &  589 & 10.5 & 89.5 \\
          & RM(M) &  858 & 20.0 & 80.0 \\
          & RM(Q) &  621 &  9.3 & 90.7 \\
\midrule
NFCorpus  & BM(M) &   73 & 34.2 & 65.8 \\
          & BM(Q) &   61 & 32.8 & 67.2 \\
          & RM(M) &   75 & 30.7 & 69.3 \\
          & RM(Q) &   59 & 32.2 & 67.8 \\
\bottomrule
\end{tabular}
\caption{Breakdown of unreachable queries ($\mathcal{W}^*_q = \emptyset$) by candidate availability. ``Neither in Top-200'' and ``$\geq$1 in Top-200'' are percentages of that row's unreachable-query count (column 3), not of all evaluation queries, indicating whether the qrels-relevant document is absent from, or present in, at least one individual retriever's top-200 list. ACORD has zero unreachable queries across all four retriever pairs (Table~\ref{tab:eval-interval-types}) and is omitted. Computed directly from the raw top-200 TREC run files for each retriever, cross-referenced against qrels.}
\label{tab:unreachable-breakdown}
\end{table*}

%% file: tables/tab-interval-width.tex
\begin{table}[!tb]
\setlength{\tabcolsep}{4pt}
\renewcommand{\arraystretch}{1.1}
\centering
\resizebox{\linewidth}{!}{%
\begin{tabular}{lcccc}
\hline
\textbf{Dataset} & \textbf{Mean} & \textbf{Median} & \textbf{Mean} & \textbf{Mean} \\
 & \textbf{Width} & \textbf{Width} & \textbf{Width} ($0.5\!\in\!\mathcal{I}_q$) & \textbf{Width} ($0.5\!\notin\!\mathcal{I}_q$) \\
\hline
MS~MARCO             & .339 & .220 & .699 & .148 \\
NQ                   & .431 & .340 & .759 & .181 \\
ACORD                & .074 & .030 & .143 & .063 \\
NFCorpus             & .270 & .075 & .780 & .088 \\
\hline
\end{tabular}}
\caption{Width of the primary optimal interval $\mathcal{I}_q = [l_q, r_q]$ (Section~\ref{subsec:optimal_weight_sets}) over reachable queries ($\mathcal{W}^*_q \neq \emptyset$), aggregated across the four retriever pairs per dataset; single-point queries contribute width 0. Splitting by whether $0.5 \in \mathcal{I}_q$ shows the gap directly: queries where standard RRF already lands inside $\mathcal{I}_q$ have intervals 4--11$\times$ wider on average than queries where it does not -- exactly the queries a predictor must move away from $w{=}0.5$ to win on are also the ones with the narrowest, hardest-to-hit targets.}
\label{tab:interval-width}
\end{table}

%% file: tables/tab-recovery.tex
\begin{table*}[!t]
\setlength{\tabcolsep}{3pt}
\renewcommand{\arraystretch}{1.1}
\centering
\resizebox{\textwidth}{!}{%
\begin{tabular}{l|cccc|cccc|cccc|cccc}
\hline
\textbf{Method} &
\multicolumn{4}{c|}{\textbf{MS~MARCO}} &
\multicolumn{4}{c|}{\textbf{NQ}} &
\multicolumn{4}{c|}{\textbf{ACORD}} &
\multicolumn{4}{c}{\textbf{NFCorpus}} \\
\cline{2-17}
 & \textbf{BM(M)} & \textbf{BM(Q)} & \textbf{RM(M)} & \textbf{RM(Q)}
 & \textbf{BM(M)} & \textbf{BM(Q)} & \textbf{RM(M)} & \textbf{RM(Q)}
 & \textbf{BM(M)} & \textbf{BM(Q)} & \textbf{RM(M)} & \textbf{RM(Q)}
 & \textbf{BM(M)} & \textbf{BM(Q)} & \textbf{RM(M)} & \textbf{RM(Q)} \\
\hline
Mean Optimal Weight         & 5.6  & 13.7 & 8.7  & 17.9
                            & 1.9  & 16.0 & 18.2 & 30.2
                            & $-$1.1 & $-$1.2 & 7.1  & 0.4
                            & $-$5.3 & 15.1 & $-$1.0 & 7.9 \\
Ridge Regressor (MBW)       & 6.5  & 15.1 & 8.9  & 18.3
                            & 3.0  & 15.8 & 17.2 & 29.8
                            & 8.2  & $-$1.5 & 4.8  & 2.2
                            & $-$3.9 & 15.2 & 0.0  & 7.4 \\
RoBERTa Regression (IW)     & 9.0  & 17.9 & 12.5 & 19.5
                            & 5.7  & 16.6 & 18.5 & 32.9
                            & $-$8.9 & 19.1 & $-$1.4 & 12.5
                            & $-$2.1 & 16.4 & 0.3  & 7.0 \\
ModernBERT (IW)             & \textbf{12.0} & 20.9 & 13.8 & 22.9
                            & 6.1  & 20.3 & 18.6 & 37.1
                            & $-$10.3 & \textbf{21.4} & $-$1.3 & 13.9
                            & $-$2.3 & 17.7 & \textbf{6.4} & 10.6 \\
ModernBERT Passage Cond.    & 11.8 & 21.2 & 13.0 & 27.2
                            & \textbf{18.8} & 22.0 & 22.8 & 38.6
                            & \textbf{15.5} & 20.0 & 11.9 & \textbf{18.4}
                            & $-$4.6 & \textbf{29.3} & 5.4  & 8.5 \\
DAT Qwen3                   & 3.0  & 6.1  & 0.5  & 6.5
                            & 7.4  & 15.6 & 11.3 & 20.1
                            & $-$2.3 & 3.1  & \textbf{23.7} & 14.8
                            & $-$10.3 & 21.4 & $-$2.1 & 15.6 \\
DAT Ministral3               & 6.0  & 15.0 & 6.8  & 18.1
                            & 14.5 & 25.7 & 28.6 & 40.0
                            & $-$8.8 & 10.0 & 7.0  & 8.9
                            & $-$25.5 & 12.3 & $-$24.9 & \textbf{16.5} \\
Qwen3 FS (IW)               & 11.5 & 26.2 & 12.5 & 28.9
                            & 9.0  & 26.6 & 26.0 & 45.2
                            & 7.2  & 5.7  & 0.0  & $-$0.5
                            & \textbf{$-$0.1} & 23.7 & 0.3  & 12.8 \\
Ministral3 FS (IW)          & 10.9 & \textbf{35.4} & \textbf{15.1} & \textbf{39.0}
                            & 5.1  & \textbf{35.0} & \textbf{39.1} & \textbf{61.1}
                            & 10.0 & 15.4 & 7.1  & 15.7
                            & $-$3.1 & 24.9 & $-$12.3 & 3.9 \\
\hline
\end{tabular}}
\caption{Headroom recovery (\%): $(\text{Method} - \text{RRF}) / (\text{Oracle} - \text{RRF}) \times 100$. Standard RRF is 0\% and the oracle is 100\% by construction. Values below zero indicate underperformance relative to standard RRF. Best per column in \textbf{bold}. (Recovery ratios are derived quantities, not directly significance-tested; see Table~\ref{tab:performance-comparison-full} for per-cell significance on the underlying metric.)}
\label{tab:recovery}
\end{table*}

%% file: tables/tab-mtc.tex
\begin{table}[!tb]
\setlength{\tabcolsep}{4pt}
\renewcommand{\arraystretch}{1.1}
\centering
\resizebox{\linewidth}{!}{%
\begin{tabular}{l|ccc}
\hline
\textbf{Baseline} & \textbf{sig.\ ($p{<}.05$)} & \textbf{Bonferroni} & \textbf{BH-FDR} \\
\hline
vs.\ RRF (standard) & 75 / 128 & 59 / 128 & \textbf{74} / 128 \\
vs.\ Mean Optimal Weight & 48 / 128 & 35 / 128 & \textbf{46} / 128 \\
\hline
\end{tabular}}
\caption{Multiple-testing summary for the query-adaptive-methods-vs-baseline family (8 methods $\times$ 16 dataset-retriever-pair combinations $=$ 128 paired $t$-tests per baseline). A cell counts if the method's mean is significantly higher than the baseline's (two-sided). Bonferroni $\alpha = 0.05/128 \approx 0.00039$; BH-FDR at $q \le 0.05$.}
\label{tab:mtc}
\end{table}

%% file: tables/tab-grid-sensitivity.tex
\begin{table}[!tb]
\centering
\small
\begin{tabular}{lccc}
\toprule
\textbf{Dataset} & \textbf{Interval $\Delta$len} & \textbf{Disagreement} & \textbf{$\Delta$Metric} \\
\midrule
MS~M. & 0.0040--0.0046 & 0.0059--0.0071 & $\le$0.0005 \\
NQ       & 0.0033--0.0039 & 0.0043--0.0050 & $\le$0.0004 \\
NFCorpus & 0.0035--0.0040 & 0.0061--0.0070 & $\le$0.0003 \\
ACORD    & 0.0025--0.0068 & 0.0087--0.0109 & $\le$0.0007 \\
\bottomrule
\end{tabular}
\caption{Sensitivity of $\mathcal{W}^*_q$ to grid resolution (0.01-step vs.\ 0.001-step),
ranges across the four retriever pairs per dataset, mean per-query values.}
\label{tab:grid-sensitivity}
\end{table}

%% file: tables/tab-error-analysis.tex

\begin{table}[!tb]
\centering
\small
\resizebox{\linewidth}{!}{%
\begin{tabular}{lccc}
\toprule
\textbf{Dataset} & \textbf{Strongest feature} & \textbf{$r$} & \textbf{Ambiguity $r$} \\
\midrule
MS~M. & query length & $+$0.07 & $-$0.01 \\
NQ       & rare/singleton terms & $+$0.09 & $-$0.02 \\
NFCorpus & term rarity (opp.\ sign) & $-$0.17 & $+$0.05 \\
ACORD    & query length\textsuperscript{1} & $+$0.28 & $+$0.08 \\
\bottomrule
\end{tabular}%
}
\caption{Point-biserial correlation between query features and misprediction (well- vs.\ poorly-predicted queries), strongest feature per dataset. \textsuperscript{1}ACORD has only 57 eval queries; its larger coefficient is high-variance.}
\label{tab:error-analysis}
\end{table}

%% file: tables/tab-k3-oracle.tex
\begin{table}[!tb]
\centering
\small
\setlength{\tabcolsep}{5pt}
\renewcommand{\arraystretch}{1.1}
\begin{tabular}{l ccc}
\toprule
\textbf{Dataset} & \textbf{Best $k{=}2$} & \textbf{$k{=}3$} & \textbf{$\Delta$} \\
\midrule
MS~MARCO  & .457 & \textbf{.477} & $+$.020 \\
NQ        & .549 & \textbf{.578} & $+$.029 \\
ACORD     & .231 & \textbf{.244} & $+$.013 \\
NFCorpus  & .473 & \textbf{.482} & $+$.009 \\
\bottomrule
\end{tabular}
\caption{Oracle ceiling: best 2-retriever pair vs.\ the 3-retriever (BM25, RM3, Qwen3) simplex oracle, per-query best weight, averaged over the eval set. Metric is MRR@10 (MS~MARCO, NQ) / nDCG@10 (ACORD, NFCorpus).}
\label{tab:k3-oracle}
\end{table}

%% file: tables/tab-k3.tex
\begin{table}[!tb]
\centering
\setlength{\tabcolsep}{12pt} 
\renewcommand{\arraystretch}{1.1}
\centering
\resizebox{\linewidth}{!}{%
\begin{tabular}{l|cccc}
\hline
\textbf{Method} & \textbf{MS~M.} & \textbf{NQ} & \textbf{ACORD} & \textbf{NFC.} \\
\hline
Oracle              & .477 & .578 & .244 & .482 \\
Headroom (Or.$-$RRF) & .230 & .264 & .112 & .098 \\
Standard RRF ($k{=}3$) & .247 & .314 & .132 & .384 \\
Ministral3 FS (IW) & .340$^{*}$ & .435$^{*}$ & .148 & .398$^{*}$ \\
ModernBERT Passage Cond. & .306$^{*}$ & .401$^{*}$ & .142 & .398$^{*}$ \\
\hline
\multicolumn{5}{l}{Recovery (\%)} \\
Ministral3 FS (IW) & \textbf{40.5} & \textbf{46.0} & \textbf{14.5} & 14.2 \\
ModernBERT Passage Cond. & 25.7 & 33.0 & 8.9 & \textbf{14.3} \\
\hline
\end{tabular}}
\caption{$k{=}3$ fusion (BM25, RM3, Qwen3): performance and headroom recovery (\%) vs.\ unweighted $k{=}3$ RRF. Stars: paired $t$-test vs.\ $k{=}3$ RRF, $^{*}p{<}.05$.}
\label{tab:k3}
\end{table}